\UseRawInputEncoding
\documentclass[a4paper,10pt]{article}
\usepackage{icad2022,amsmath,epsfig,times,url,hyperref}

\usepackage{listings}
\usepackage{xcolor}

\hypersetup{
    colorlinks,
    citecolor=darkgray,
    linkcolor=darkgray,
    urlcolor=darkgray
}

\lstdefinelanguage{args}{
sensitive=false,
alsoletter={.},
moredelim=[s][\color{red}]{<}{>},
moredelim=[s][\color{blue}]{[}{]},
moredelim=[is][\color{orange}]{:}{:},
keywords=[10]{...},
keywordstyle=[10]{\color{magenta}},
}
\lstnewenvironment{bash}
{\lstset{numbers=left,language=bash,keywordstyle={\color{blue}}}}
{}

\title{T\=oky\=o Kion-On: Query-Based Generative Sonification of Atmospheric Data}

\name{Stefano Kalonaris} 
\address{RIKEN AIP \\ 
Music Information Intelligence Team \\ 
{\tt stefano.kalonaris@riken.jp}} 
\begin{document}
\ninept
\maketitle

\begin{sloppy}
\begin{abstract}
Amid growing environmental concerns, interactive displays of data constitute an important tool for exploring and understanding  the impact of climate change on the planet's ecosystemic integrity. 
This paper presents {\em T\=oky\=o kion-on}, a query-based sonification model of Tokyo's air temperature from 1876 to 2021. The system uses a recurrent neural network architecture known as LSTM with attention trained on a small dataset of Japanese melodies and conditioned upon said atmospheric data. After describing the model's implementation, a brief comparative illustration of the musical results is presented, along with a discussion on how the exposed hyper-parameters can promote active and non-linear exploration of the data. 
\end{abstract}

\section{Introduction}
\label{sec:intro}
The planet's rising temperature is an indisputable trend, with catastrophic effects for the ecosystemic balance and all life on Earth as we know it.
With growing consensus and awareness that a tipping point of this fragile dynamical, complex, and adaptive system has perhaps already been reached (or that is inevitably close), what has been dubbed as the ``climate crisis" is polarizing opinions, actions, and stakeholders' interest across the globe. It has been argued \cite{10.1175/BAMS-D-15-00223.1} that alternative ways to convey  environmental data can include auditory display methods, which can help overcome the issues related to engagement with graphical representations. 
However, in cases where the data is not sufficiently multidimensional and/or complex to require  alternative means of display and insight, mapping methods which explicitly couple the range of data variables to that of musical parameters (e.g., pitch, dynamics, etc.) risk offering an equally predictable and unengaging representation of the phenomena under consideration. While general consensus argues that ``mappings will follow simple intuitive basics, e.g., temperature to pitch, location to spatial rendering, etc., and the sound design shall evoke climate associations that are straightforward, e.g., as known from weather conditions" \cite[p.11]{10.4018/978-1-4666-6228-5.ch001}, this can be problematic in the case of temperature data, which follows a linear growth trend; the aftermath of a few degrees in the planet's temperature has irrevocable consequences, but to appropriately convey this via audiovisual display almost necessarily involves re-scaling the temperature range so that such a modest difference value can be appropriately perceived and communicate the sense of urgency that it warrants. This, in turn, might result in unwanted sound caricatures or contour exaggerations, to the detriment of musicality. 

This paper, on the other hand, foregrounds aesthetic concerns and presents a query-based model of auditory display where a more subtle correspondence is sought. To this end, mapping of the data is applied to a neural network's internal representation of the musical corpus' parametric encoding. Given the {\em black box} nature of neural networks, the sonification model confers a degree of opacity and surrogacy \cite{KalonarisZannos2021} which might elicit meaning-making processes rather than meaning-finding or meaning-carrying. 

Furthermore, the environmental data of interest constitutes but the initial inspiration behind the adoption of the author's chosen sonification design and procedure. Along the continuum between concrete to abstract sonification as outlined in \cite{VickersHogg2006}, {\em T\=oky\=o kion-on} is mostly focusing on the latter, and should be intended primarily as a generative task rather than an empirically-based sonification. In accordance to more current trends in the broader field of aesthetics, where the sensibility has continued to move towards a pragmatist stance, the author establishes affinity with a notion of interpretation that welcomes and embraces the role of indeterminacy \cite{ShustermanInterview}, fully aware that 
``by casting the definition in terms of science and objectivity alone, the more complex interrelationship between the arts and using data as a direct determinate is lost" \cite[p.208]{Gresham-Lancaster2012}.
While sequential/chronological auditory display of the data is certainly possible, much like the majority of environmental data sonification systems that focus on the temporal evolution of the phenomenon of interest, the author's model affords a non-linear approach, allowing the user to single out years of interest, change the length of the generated melody, experimenting with different priming seeds, and, ultimately, procedurally combine results for compositional or purely explorative goals.

\section{Related Work}
\label{sec:related_work}
Different methods are normally used to auditorily display data. These include \emph{Earcons} or \emph{Auditory Icons} (sound aliases for events or actions, typically used as auditory aids), \emph{Audification} \cite{Dombois01usingaudification} (direct sonification of data as a series of sound pressure values, using re-sampling and filtering), \emph{Parameter-Mapping Sonification} (PMSon) \cite{Worrall2019PMS} (mapping and transfer functions reveal latent structures in the data), \emph{Model-Based Sonification} (MBSon) \cite{10.1109/JPROC.2004.825904} (a virtual instrument model built from the data of interest is played via user interaction), and \emph{Wave Space Sonification} (WSSon) \cite{Hermann2018} (a data-driven trajectory scans a scalar field).

Sonification of environmental and atmospheric data is certainly not a novel endeavor, and the above methods have been explored in \cite{Quinn2001,Flowers2001,Hermann2003,Polli2004}. 
An accomplished example of artistic-led sonification of weather data which manages to integrate disparate concerns and viewpoints is the {\em Locus Wrath} \cite{10.1162/leon_a_01339}, a multi-modal interactive system for dance performance, where aesthetic considerations and high ecological validity are coupled with the concurrent usage of different sonification methods.

In the realm of auditory display, the use of artificial neural networks (ANNs) is still relatively a novelty, unlike in other subfields of music and sound computing where these are increasingly becoming the norm. For the most part, precedents in this direction tend to deal with data of interest other than climate-related. Notwithstanding, the recent work of Herrmann \cite{pmlr-v123-herrmann20a} is an important reference when reviewing deep learning-based sonifications. This approach is followed in \cite{Halac2021} and there, too, the sonification process auditorily displays inner layers of the network's activation functions activity. 

One example of ANNs application to atmospheric data is documented in \cite{Rodney2021}, where a variational auto-encoder (VAE) model released by Google's Magenta team is used to sonify smart city data including weather data. The VAE is employed to blend a source and a target musical motif with a morphing coefficient proportional to the change in rainfall. In the same work, temperature data and wind speed are also mapped to an LFO's fundamental frequency and amplitude. This work is perhaps the most similar to the system presented in this paper, so far as the use of ANNs is concerned. 

{\em T\=oky\=o kion-on} uses a recurrent neural network (RNN) known as long short-term memory (LSTM). This architecture has been the most widely used in tasks involving temporal dependencies (e.g., time series) before the advent of newer models using self-attention \cite{10.5555/3295222.3295349}, with the music domain not being an exception in this regard \cite{huang2018music, ChengZhi2018MusicTransformer,HawthorneMLCD2018}. However, while there are countless examples of generative music systems built using LSTMs \cite{performance-rnn-2017,Liang2017AutomaticSC,Sturm1260836,DBLP:journals/corr/abs-1811-08045}, this approach remains under-explored in sonification tasks.

\section{Model}
\label{sec:model}
{\em T\=oky\=o kion-on} can be considered both a MBSon and a PMSon model in the currently accepted taxonomy of auditory display methods. On one hand it is a model where the data of interest is explored via means of user interaction. However, the data model is not the atmospheric data itself but the representation of a musical corpus learned by an ANN. On the other hand, an arbitrary mapping between the data and sound parameters is established so that the sonic output is, virtually, a surrogate of the phenomena under consideration. In the form described in this paper, and as of the time of writing, {\em T\=oky\=o kion-on} is available online at \url{https://gitlab.com/skalo/tokyokionon}. 
Before delving into the details, a brief description of both the training musical corpus and the atmospheric data is given.

\subsection{Data}
\label{subsec:data}
The neural network was trained on a small dataset of public domain Japanese melodies in .musicxml format hosted online,\footnote{\url{http://www.daisyfield.com/music/htm/-genres/japan.htm/}} excluding those arranged for piano, and all melodies were ``normalized" to the same key (it is possible, if desired, to re-train the model with the dataset augmented to all 12 keys).

Atmospheric data was obtained from the Japan Meteorological Agency's website,\footnote{\url{https://www.data.jma.go.jp}} using the monthly mean daily maximum temperature \href{https://www.data.jma.go.jp/obd/stats/etrn/view/monthly_s3_en.php?block_no=47662&view=2}{table}.
The specific architecture of ANN (see Section \ref{subsec:architecture}) used allows to control the randomness of predictions with a hyper-parameter which is, incidentally, known as {\em temperature} (one must be careful to disambiguate this from the atmospheric data homonymous measure). The atmospheric data was pre-processed to obtain two year-indexed normalized vectors that serve as the ANN's temperature hyper-parameter for a given year's sonification query. One of these vectors is used for the randomness control of the pitch predictions, the other for the randomness control of the notes' duration (see Section \ref{subsec:generation} for a detailed explanation). 

The pitch temperature vector is obtained by calculating the cosine similarity between a given year's forward difference of the monthly air temperature average and the corresponding forward difference for the reference year's (i.e., 1876) and subtracting the resulting value from 1. This is because cosine similarity returns a value between $[0,1]$ where $1$ is the identity but, for the ANN's temperature hyper-parameter, $1$ should correspond to the highest probability of sampling from unlikely candidates in the prediction (see Section \ref{subsec:generation}).
As for the duration temperature, the normalized vector of the annual mean daily maximum temperature is used.

\subsection{Architecture \& Training}
\label{subsec:architecture}
Notwithstanding ample room for experimentation with different models, the LSTM architecture was deemed
appropriate because it is apt for tasks involving time series modeling (thus it works well both for atmospheric data and melodic sequences, alike). One known shortcoming of LSTMs is the ability to model time dependencies at meso and macro level, which in music would translate to relational structures beyond musical periods (e.g., sections, repetitions, and long-term structure, in general). To address these levels, models using self-attention such as the Transformer \cite{10.5555/3295222.3295349} are more useful. However, there are two main reasons why this was not the traveled path: 1) the corpus used in training comprises exclusively short traditional melodies, whose local dependencies are well catered for by LSTMs and 2) Transformers require vast amounts of training data which, for the musical genre of choice, is not currently available.

Just like RNNs, LSTMs are chains of connected modules called units. Unlike RNNs, an LSTM unit comprises 4 layers (three sigmoid and one hyperbolic tangent), as shown in Figure \ref{fig:lstm-unit}. {\em T\=oky\=o kion-on} uses 256 such units and employs the {\em attention} mechanism, which overcomes RNNs limitations in the encoder-decoder architecture by allowing the network to learn where to attend/focus most in the output sequence. 
{\em T\=oky\=o kion-on} takes two input sequences, one for the pitch and one for the duration values, embeds these into vectors and concatenates these into a single input vector to the recurrent layers. After training, it produces two outputs, a prediction for the notes and another for their duration.

\begin{figure}[ht]
  \centering
  \includegraphics[width=0.5\textwidth]{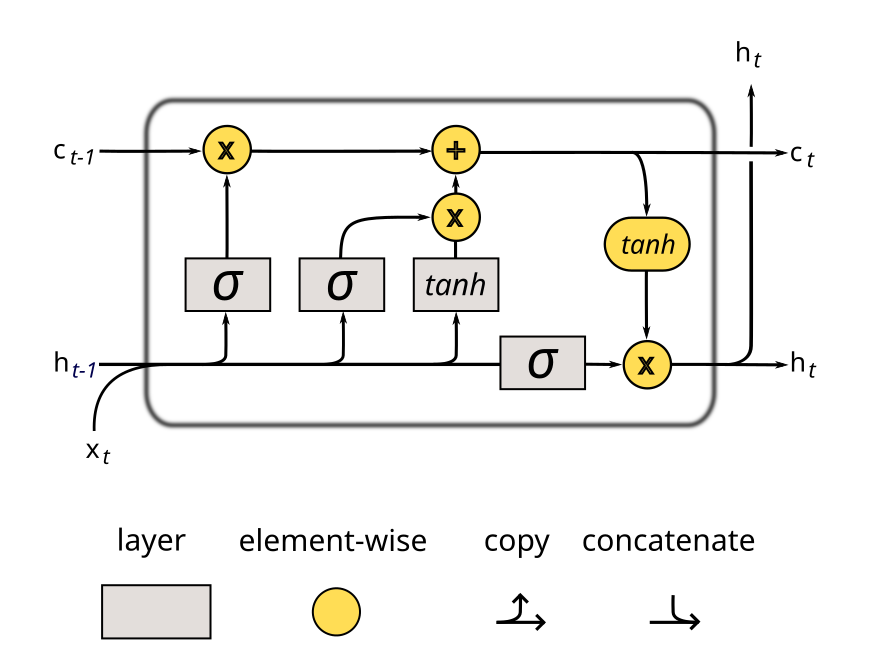}
  \caption{One unit of an LSTM network. Gray blocks are neural network activation layers, yellow ellipses are point-wise operators, and directed lines are vector transfers, copies or concatenations. $C$ is the cell state, $x$ is the input, and $h$ the output, at a given time $t$.}
  \label{fig:lstm-unit}
\end{figure}

Given the modest number of samples (46) in the training dataset, the model was trained without using GPU on a 8-core AMD processor and converged after 205 epochs with {\em Early Stopping} and {\em patience} of $10$.

\subsection{Generation}
\label{subsec:generation}
Users can query a year in the range of the atmospheric data at the time of writing (i.e., 1876 to 2021). Without any further argument, the generative part of the system ``predicts" a melody based on the weights learned during training, on the corresponding pitch and duration temperatures in the year-indexed vectors obtained during data pre-processing (see Section \ref{subsec:data}), and on default parameters that dictate the output's sequence length. A brief explanation of how temperature works in an LSTM is now due.

Neural networks use a logit vector $z$ where $z=(z_{1},\ldots,z_{n})$ and apply an activation function to produce a probability vector $q=(q_{1},\ldots,q_{n})$ over predicted output classes by comparing $z_{i}$ with the other logits. {\em Softmax} is normally the activation function of choice for the last layer of a neural network.
In LSTMs, the temperature hyper-parameter $T$ controls the randomness of the predictions, effectively being a scaling factor for the logits, when computing the softmax output. Probability $q{i}$, in this case, would then be calculated as shown in Equation \ref{eq:temp_softmax}.

\begin{equation}\label{eq:temp_softmax}
    q_{i} = \frac{\exp(\frac{z_{i}}{T})}{\sum_{j}^{n}\exp(\frac{z_{j}}{T})}
\end{equation}
\medskip

In the specific implementation of {\em T\=oky\=o kion-on}'s LSTM, $T$ is a value in the range $[0, 1]$. When equal to $1$, the softmax is computed directly on the logits, whereas for temperatures $<1$ the softmax will yield larger values, making the LSTM more confident in the prediction (since less input can activate the output layer) albeit more conservative, less prone to sample from unlikely candidates.

\section{Output Examples}
\label{sec:examples}

In terms of user interaction, and as for the design stage discussed in this paper at the time of writing, a convenience script has been included in the repository that hosts the code. This script can be executed directly by providing arguments of choice for the exposed hyper-parameters. An example of shell command is shown below, with abbreviated arguments (i.e., \texttt{-y} instead of \texttt{--year}, etc.):

\begin{lstlisting}[language=Python, showstringspaces=false, basicstyle=\ttfamily, caption= Example of invoking the sonification script via command-line.]
>>> python tokyokionon.py -y 1984  -s [[`A4'],[0.5]] -mxx 8 -mxl 16 -sql 16
\end{lstlisting}

\medskip
Using default values for the generation allows for direct comparison between different query years, since the priming seed is kept invariant (pitch: $A4$, duration in quarter length: $1.0$) and the output sequence length is set to $16$ tokens. Figure \ref{fig:compare_melodies} shows the generated melodies for the reference year (1876) and for one of the years that are most dissimilar to it according to the monthly forward difference cosine similarity metric (see Section \ref{subsec:data}), which affects the note randomness. As one can verify, the notes vary, whereas the duration pattern is the same. In fact, 1980's duration temperature value stands at $0.37$, a value perhaps not sufficiently large to trigger variations in the notes' duration values.

\begin{figure}[ht]
    \begin{minipage}[h]{0.48\textwidth}
      \centering
      \includegraphics[width=\textwidth]{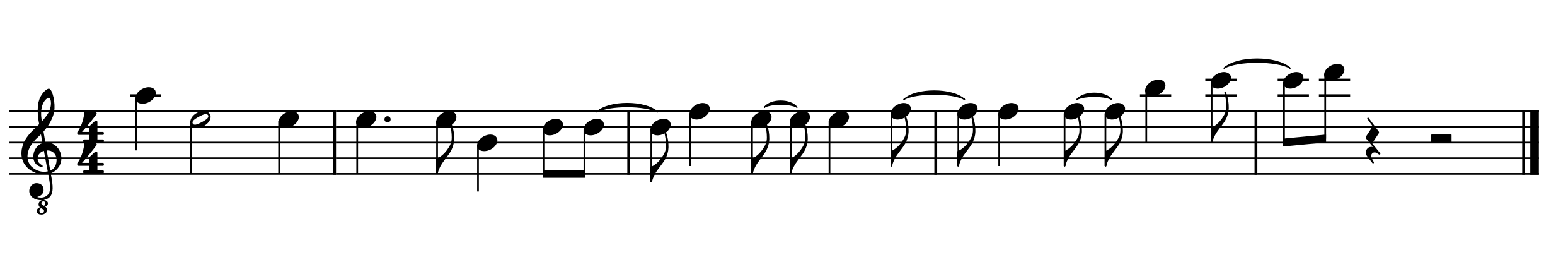}
      \centerline{(a) 1876}\medskip
    \end{minipage}
    \begin{minipage}[h]{0.48\textwidth}
      \centering
      \includegraphics[width=\textwidth]{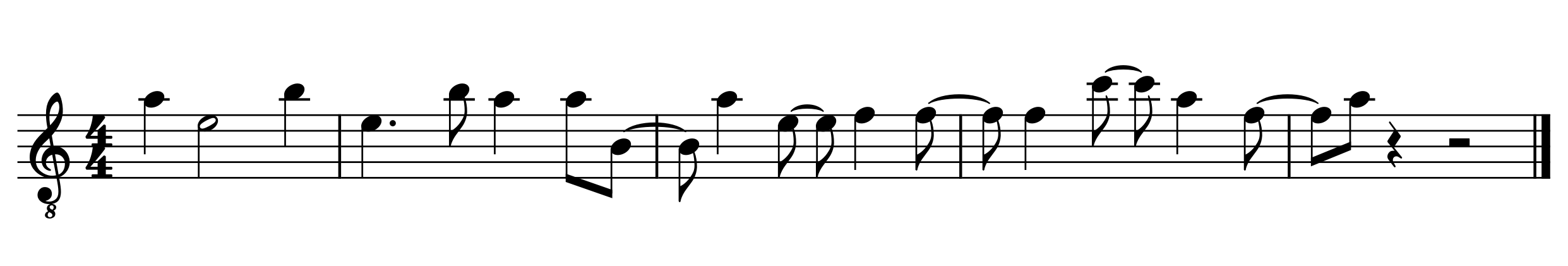}
      \centerline{(b) 1980}\medskip
    \end{minipage}
\caption{Comparing melodies generated for 1876 and 1980 using default parameter values.}
\label{fig:compare_melodies}
\end{figure}

In addition to the year query, the user has access to additional parameters. For example, querying 2021 with an empty seed yields the melody shown in Figure \ref{fig:tokyokionon_2021}. This time, the duration values are considerably different from the reference year, since the normalized value of the annual mean temperature for 2021 is the maximum attainable value, $1$.

\begin{figure}[ht]
  \centering
  \includegraphics[width=0.48\textwidth]{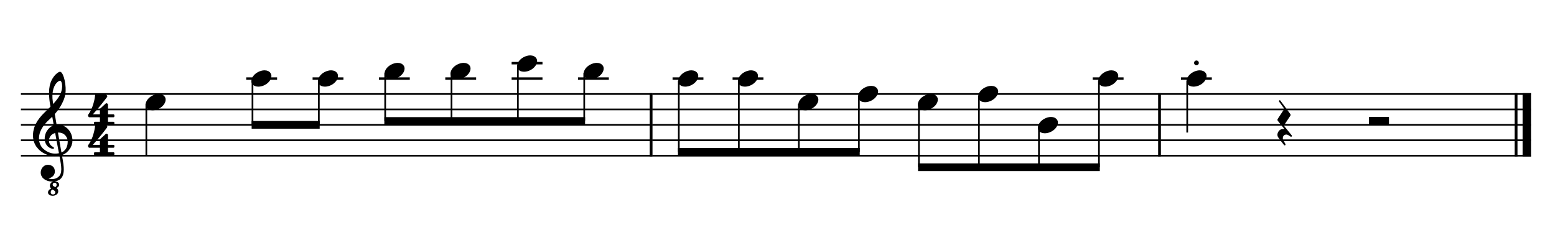}
  \caption{Melody generated for data points corresponding to 2021, without a priming seed.}
\label{fig:tokyokionon_2021}
\end{figure}

Figure \ref{fig:tokyokionon_2004}, instead, shows a query for 2004, the second warmest with a duration temperature of $0.826$, as well as a relatively high note temperature of $0.761$. In this instance, the query parameter for maximum extra notes was increased to 32 and the priming seed was reverted to the default, for comparison with the melodies shown in Figure \ref{fig:compare_melodies}.

\begin{figure}[ht]
  \centering
  \includegraphics[width=0.48\textwidth]{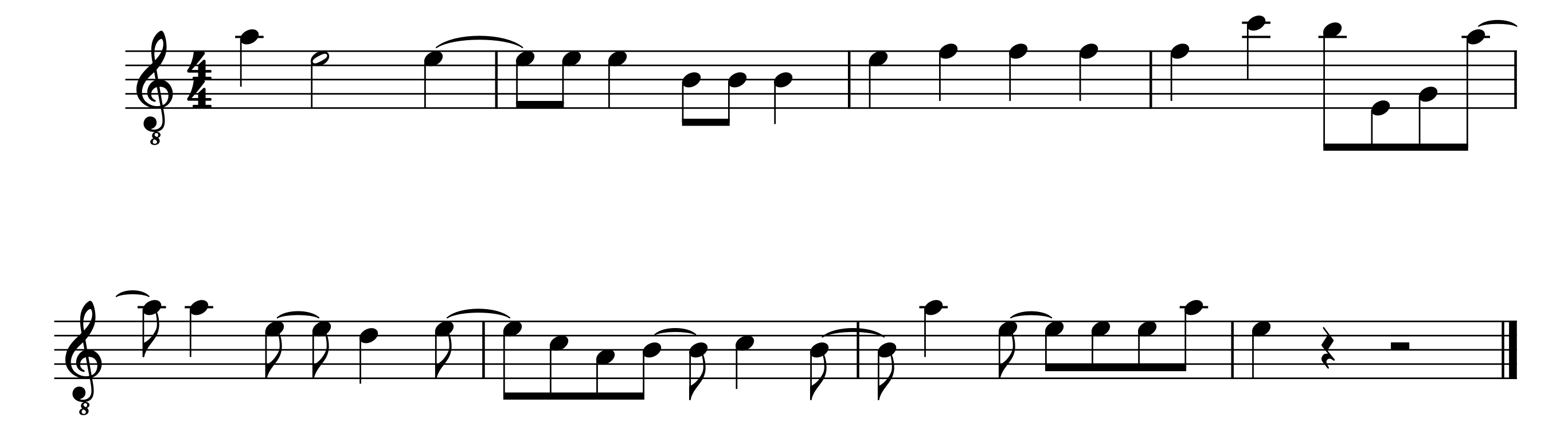}
  \caption{Melody generated for data points corresponding to 2004, with default priming seed and maximum extra notes length of 32.}
\label{fig:tokyokionon_2004}
\end{figure}

As expected, both the notes and their duration differ from the 1876 melody.
It is also possible to inspect both the attention matrix and the note candidates distribution of a generated melody, for added insight. The former shows how the neural network attends every hidden state from each encoder node at every time step, determining which of these are considered more informative for making predictions. The latter portrays the likelihood of notes being selected over time.
Examples of these representations, which are visually akin to heat maps, are shown in Figures \ref{fig:att-matrix} and \ref{fig:preds_distro}.

\begin{figure}[ht]
  \centering
  \includegraphics[width=0.45\textwidth, trim={1.5cm 0 0 0},clip]{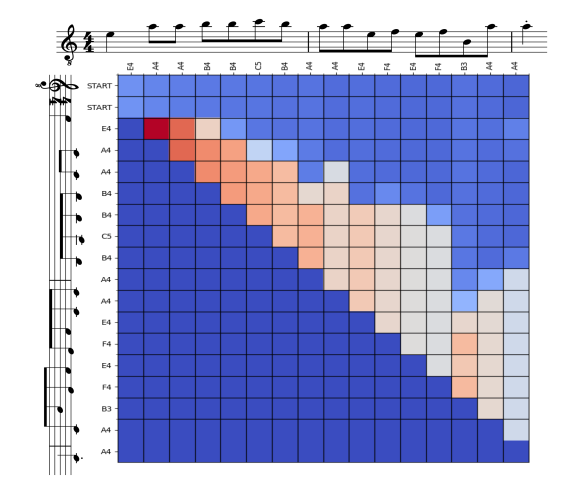}
  \caption{Attention matrix for the 2021 example. The amount of ``redness" of a cell is proportional to the attention given to the hidden state of the model corresponding to the $y$ position, when predicting the note on the $x$ position.}
\label{fig:att-matrix}
\end{figure}

\begin{figure}[ht]
  \centering
  \includegraphics[width=0.5\textwidth]{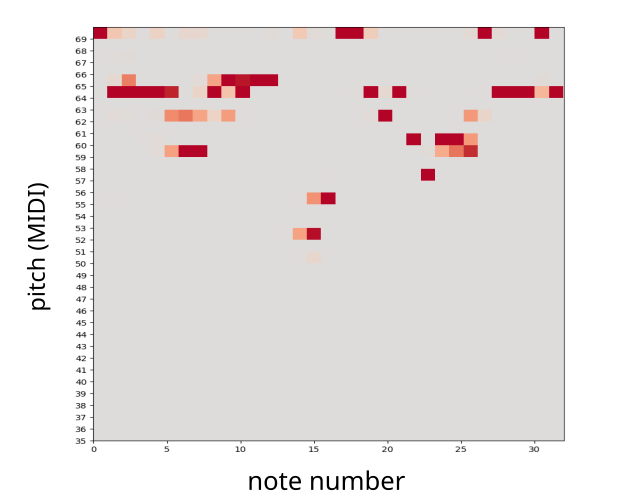}
  \caption{The distribution of the note candidates for the 2004 example. The darker (red) is a note at any given time step, the more likely it is that the model will select it.}
\label{fig:preds_distro}
\end{figure}

If, instead of single-year queries, one wishes to obtain a sonification of given range of years in the atmospheric dataset, this could be easily done by, for example, concatenating melodies for all desired entries. One would perhaps use a shorter sequence length, as demonstrated in Figure \ref{fig:range_melodies} which shows the melodies for the first and the last ten years in the dataset (1876-1886 and 2011-2021) generated using an empty seed and a sequence length of $4$.

\begin{figure}[ht!]
    \begin{minipage}[h]{0.48\textwidth}
      \centering
      \includegraphics[width=\textwidth]{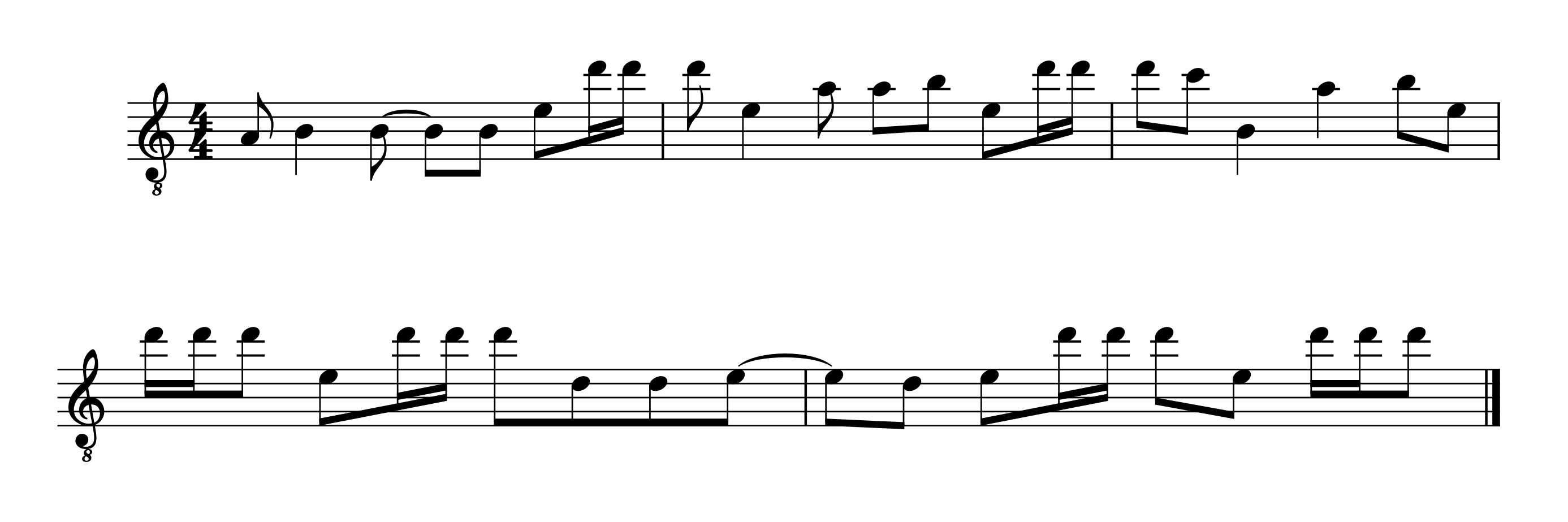}
      \centerline{(a) 1876-1886}\medskip
    \end{minipage}
    \begin{minipage}[h]{0.48\textwidth}
      \centering
      \includegraphics[width=\textwidth]{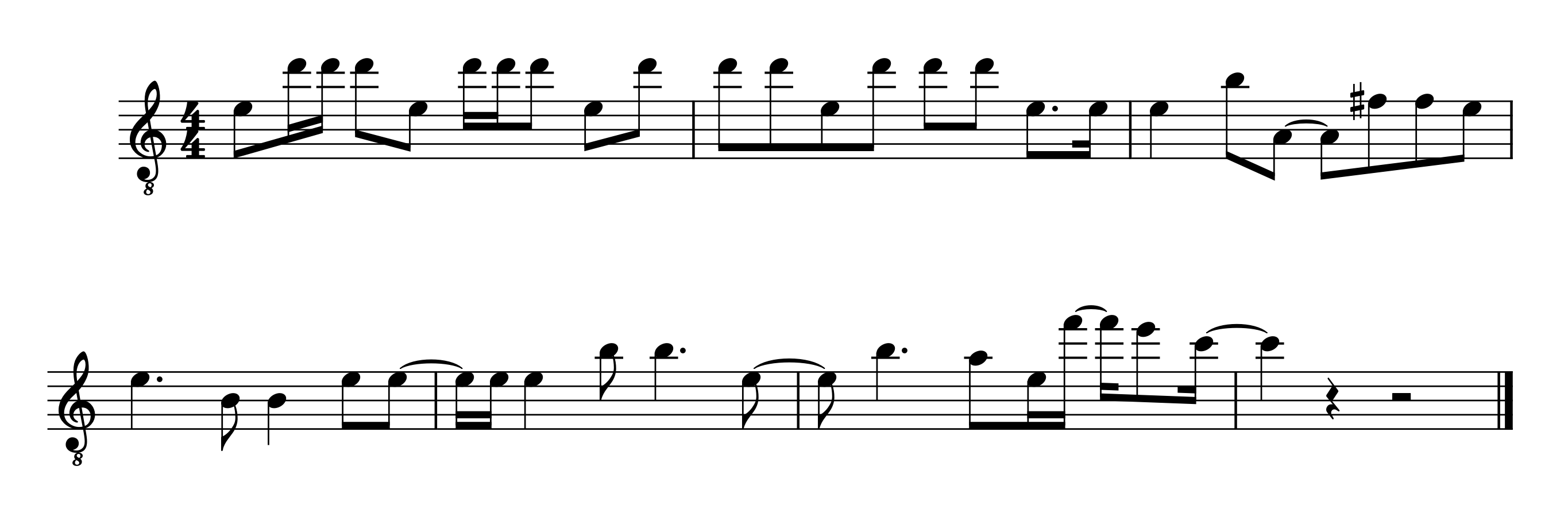}
      \centerline{(b) 2011-2021}\medskip
    \end{minipage}
\caption{First and last ten-year range in the air temperature dataset, sonified using a sequence length of $4$ and an empty priming seed.}
\label{fig:range_melodies}
\end{figure}

\section{Discussion}
\label{sec:discussion}
This paper presented {\em T\=oky\=o kion-on}, an interactive sonification model of Tokyo's air temperature from 1876 to 2021 based on a popular recurrent neural network architecture, and trained on a dataset of Japanese melodies. 
{\em T\=oky\=o kion-on} as presented here is still at a prototypical stage, and a formal evaluation has yet to be carried out. However, in the author's informal experimentation with the model's hyper-parameters, this system provided an interesting tool for querying environmental data and melodic generation alike.
By exploiting the intrinsic affordance of the network to control the randomness of the output candidates, {\em T\=oky\=o kion-on} offers a simple yet effective interface to explore both single year entries as well as entire ranges of years, and to easily compare different years auditorily. 
Unlike the leading trend in sonification, whereby the phenomenon under consideration is rendered sequentially so as to display its development and temporal unfolding in linear time, this work allows non-linear exploration of the data of interest, by querying specific years of choice. It is of course possible if trivial to simply concatenate all years' generated melodies in succession to obtain a more mainstream rendition, if so desired.
Furthermore,  {\em T\=oky\=o kion-on} can be used creatively and beyond the sonification task, to generate sequences (melodies) of arbitrary length primed with different seeds (providing these exist in the learned vocabulary). 

The system could be improved greatly if trained on a larger corpus, but this was hindered by the endemic lack of Japanese music datasets. More experimentation should be carried out with different ANNs models and it would be valuable to implement an interactive web-based version of the system. Moreover, it would be interesting to expand the scope of the system to include different pre-trained models (on different corpora) or allowing the users to upload different environmental data to condition the generative output.

\bibliographystyle{IEEEtran}
\bibliography{bibliography}

\end{sloppy}
\end{document}